\documentclass{latex/webofc}
\usepackage[varg]{txfonts}   % Web of Conferences font
\usepackage{xcolor}
\usepackage{hyperref}
\graphicspath{{figures/}}

\hypersetup{bookmarksnumbered=true, bookmarksopen=true, bookmarksopenlevel=0}
\hypersetup{pdftitle={Scalable ATLAS pMSSM computational workflows using containerised REANA reusable analysis platform}, pdfauthor={Donadoni, Feickert, Heinrich, Liu, Me{\v c}ionis, Moisieienkov, {\v S}imko, Stark, Vidal Garc\'{\i}a}}
\hypersetup{colorlinks,breaklinks}
\hypersetup{linkcolor=blue,citecolor=blue,filecolor=black,urlcolor=blue}

\begin{document}

\title{Scalable ATLAS pMSSM computational workflows using containerised REANA reusable analysis platform}

\author{\firstname{Marco} \lastname{Donadoni} \inst{1} \and
        \firstname{Matthew} \lastname{Feickert} \inst{2} \and
        \firstname{Lukas} \lastname{Heinrich} \inst{3} \and
        \firstname{Yang} \lastname{Liu} \inst{4} \and
        \firstname{Audrius} \lastname{Me\v cionis} \inst{1} \and
        \firstname{Vladyslav} \lastname{Moisieienkov} \inst{1} \and
        \firstname{Tibor} \lastname{\v Simko} \inst{1} \fnsep\thanks{Corresponding author \email{tibor.simko@cern.ch}} \and
        \firstname{Giordon} \lastname{Stark} \inst{5} \and
        \firstname{Marco} \lastname{Vidal Garc\'{\i}a} \inst{1}}

\institute{CERN, Geneva, Switzerland \and
           University of Wisconsin--Madison, Madison, United States \and
           Max-Planck-Institut f\"ur Physik, M\"unchen, Germany \and
           Sun Yat-Sen University, China \and
           SCIPP, UC Santa Cruz, United States}

\abstract{%
 In this paper we describe the development of a streamlined framework for large-scale ATLAS pMSSM reinterpretations of LHC Run-2 analyses using containerised computational workflows.
The project is looking to assess the global coverage of BSM physics and requires running O(5k) computational workflows representing pMSSM model points.
Following ATLAS Analysis Preservation policies, many analyses have been preserved as containerised Yadage workflows, and after validation were added to a curated selection for the pMSSM study.
To run the workflows at scale, we utilised the REANA reusable analysis platform. We describe how the REANA platform was enhanced to ensure the best concurrent throughput by internal service scheduling changes.
We discuss the scalability of the approach on Kubernetes clusters from 500 to 5000 cores.
Finally, we demonstrate a possibility of using additional ad-hoc public cloud infrastructure resources by running the same workflows on the Google Cloud Platform.

}
\maketitle

\section{Introduction}\label{sec:intro}

We have developed a streamlined framework for large-scale pMSSM reinterpretations of ATLAS analyses of LHC Run-2 using containerised computational workflows.
The project is looking to assess the global coverage of BSM physics and requires running numerous computational workflows representing pMSSM model points.
The framework builds upon the idea of RECAST-ing analyses~\cite{Cranmer:2010hk} and takes into account the experiences with the previous ATLAS pMSSM reinterpretations from LHC Run-1 period~\cite{ATLAS:2015wrn}.

\begin{figure}
\centering
\includegraphics[width=0.9\textwidth]{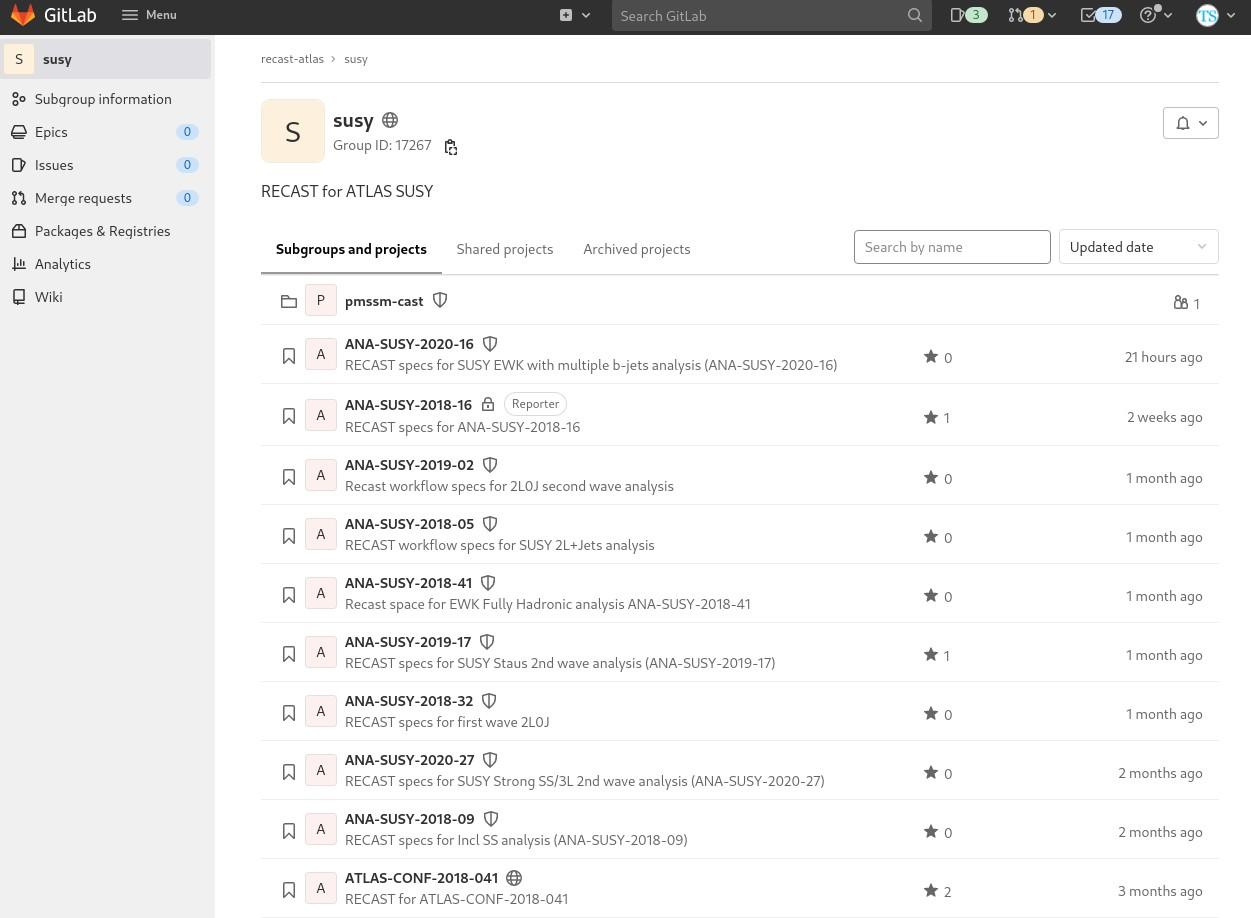}
\caption{A screenshot of the ATLAS SUSY group analyses preserved on GitLab. Each repository is labeled with the internal ATLAS analysis identifier and contains both workflow files and additional data files needed for the computational processing.}
\label{fig:pmssmgitlab}
\end{figure}

Following the ATLAS analysis preservation policies, many ATLAS analyses have been preserved as containerised Yadage workflows.
After validation they are added to a curated selection of analyses suitable for the pMSSM study.
Figure~\ref{fig:pmssmgitlab} shows one such repository for the supersymmetry searches.

\begin{figure}
\centering
\includegraphics[width=0.9\textwidth, trim=0 40px 0 0, clip]{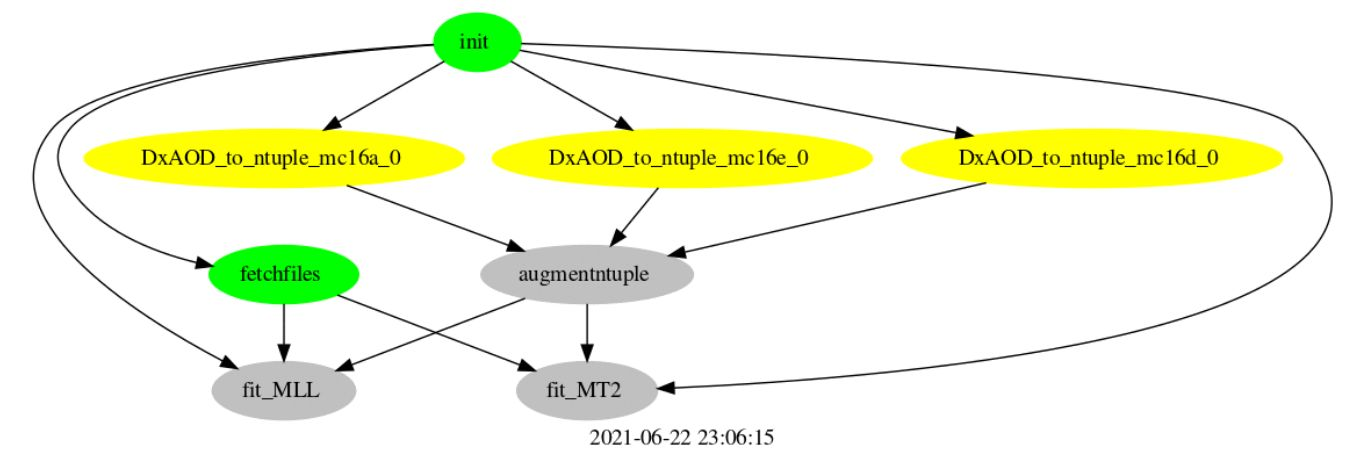}
\caption{A typical pMSSM workflow. The computational runtime is about 10 minutes without systematics (test payload) and about 10 hours with all systematics (real payload).}
\label{fig:dag}
\end{figure}

One typical pMSSM computational workflow is presented in Figure~\ref{fig:dag}.
The workflow consists of three time-consuming ntupling steps that process data files and run in parallel.
The workflow ends with a latter fitting steps that run afterwards.
The dependency of steps in the computational graph is rather simple.
The complexity of the problem lies in having to run several thousands of these workflows in order to cover a sufficient number of pMSSM model points.

It was the goal of the present work to study the feasibility of running several thousands of these containerised workflows in parallel in an automated way in order to facilitate typical pMSSM studies.

\section{Method}\label{sec:method}

The computational workflows were run at scale using the REANA reusable analysis platform~\cite{Simko:2018zzz}.
The computational backend was the Kubernetes cluster of various sizes (from 500 cores up to 5000 cores).
We have been varying several parameters of the cluster such as the number of nodes and the required memory and studied the maximum number of pMSSM workflows that the platform can handle concurrently.
After performing several such computational experiments, we have improved the scheduling efficiency of REANA to increase the running bandwidth for the pMSSM style of workflows.

\begin{figure}
\centering
\includegraphics[width=0.9\textwidth]{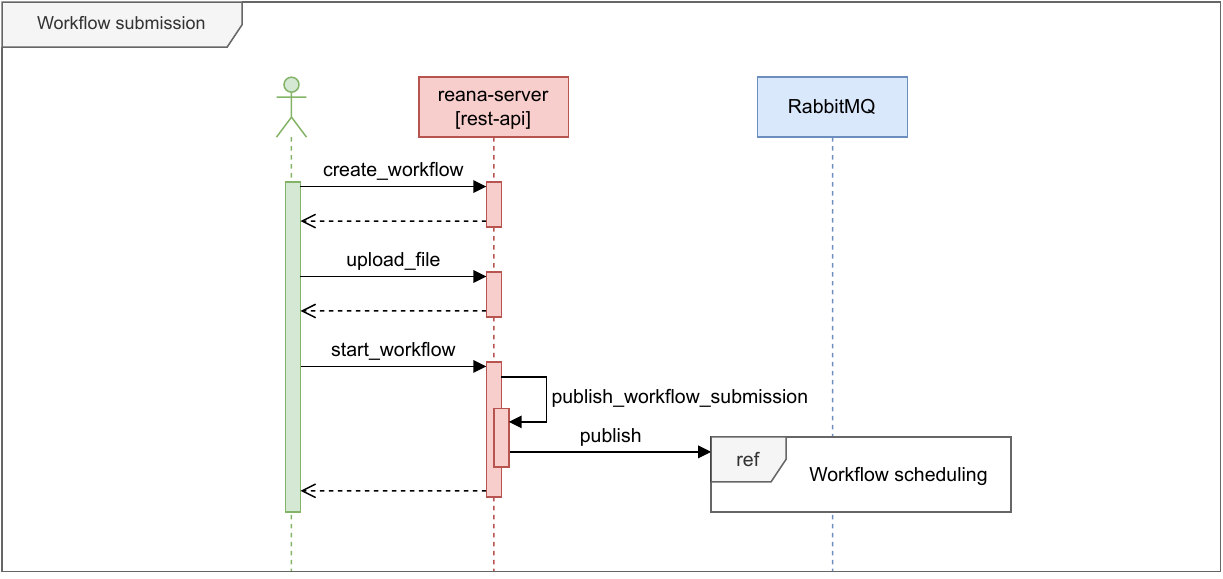}
\caption{The sequence diagram showing how REANA schedules incoming workflows after submission.
The submitted workflows are announced via message queue that is later processed by the workflow scheduler in Figure~\ref{fig:reanascheduler2}.}
\label{fig:reanascheduler1}
\end{figure}

Figure~\ref{fig:reanascheduler1} shows the sequence diagram of the workflow submission stage.
The incoming workflows are stored in a queue that is later processed by the scheduler.
The first task was to improve the performance of the REANA platform's server submission end points to allow many concurrent workflow starting requests.

\begin{figure}
\centering
\includegraphics[width=0.9\textwidth]{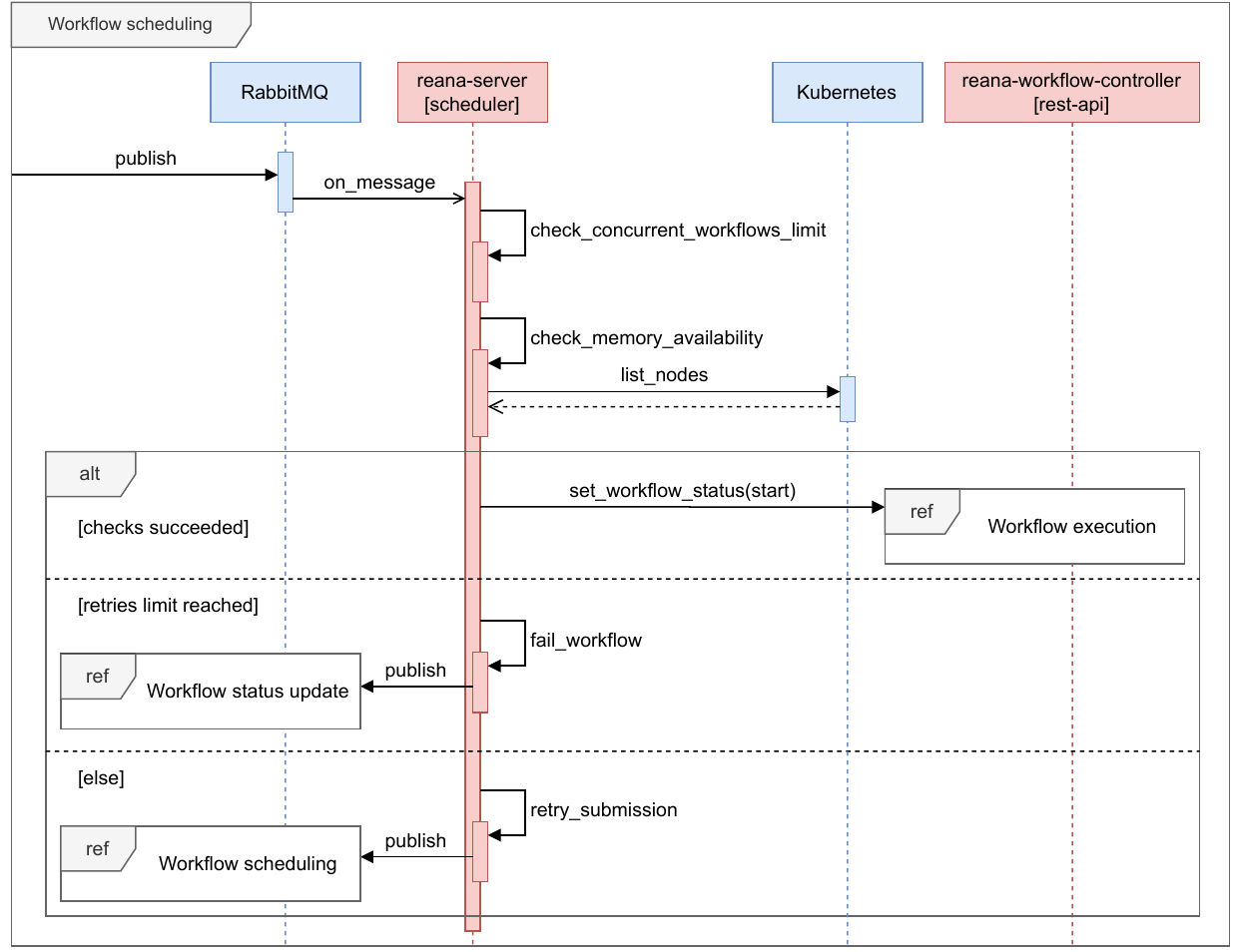}
\caption{The sequence diagram showing how REANA schedules queued workflows.
The system checks for available resources before allowing workflow runs for execution.
The checking and rescheduling workflow offers several possibilities for optimisations.
The workflows accepted for execution are further processed in Figure~\ref{fig:reanascheduler3}.}
\label{fig:reanascheduler2}
\end{figure}

Figure~\ref{fig:reanascheduler2} shows the next stage of the process, namely how the submitted workflows are being consumed from the incoming queue.
The scheduler first checks whether the incoming workflow does not exceed the limits on the total number of workflow the system could handle as well as currently available free memory on the Kubernetes cluster.
If the checks succeed, the workflow is accepted for execution.
In the opposite case the incoming workflow is being rescheduled and attempted to be accepted for execution several times whilst waiting for the Kubernetes cluster resources to liberate.
If the workflow cannot be scheduled for a substantial amount of time, a failure is declared.

\begin{figure}
\centering
\includegraphics[width=0.95\textwidth]{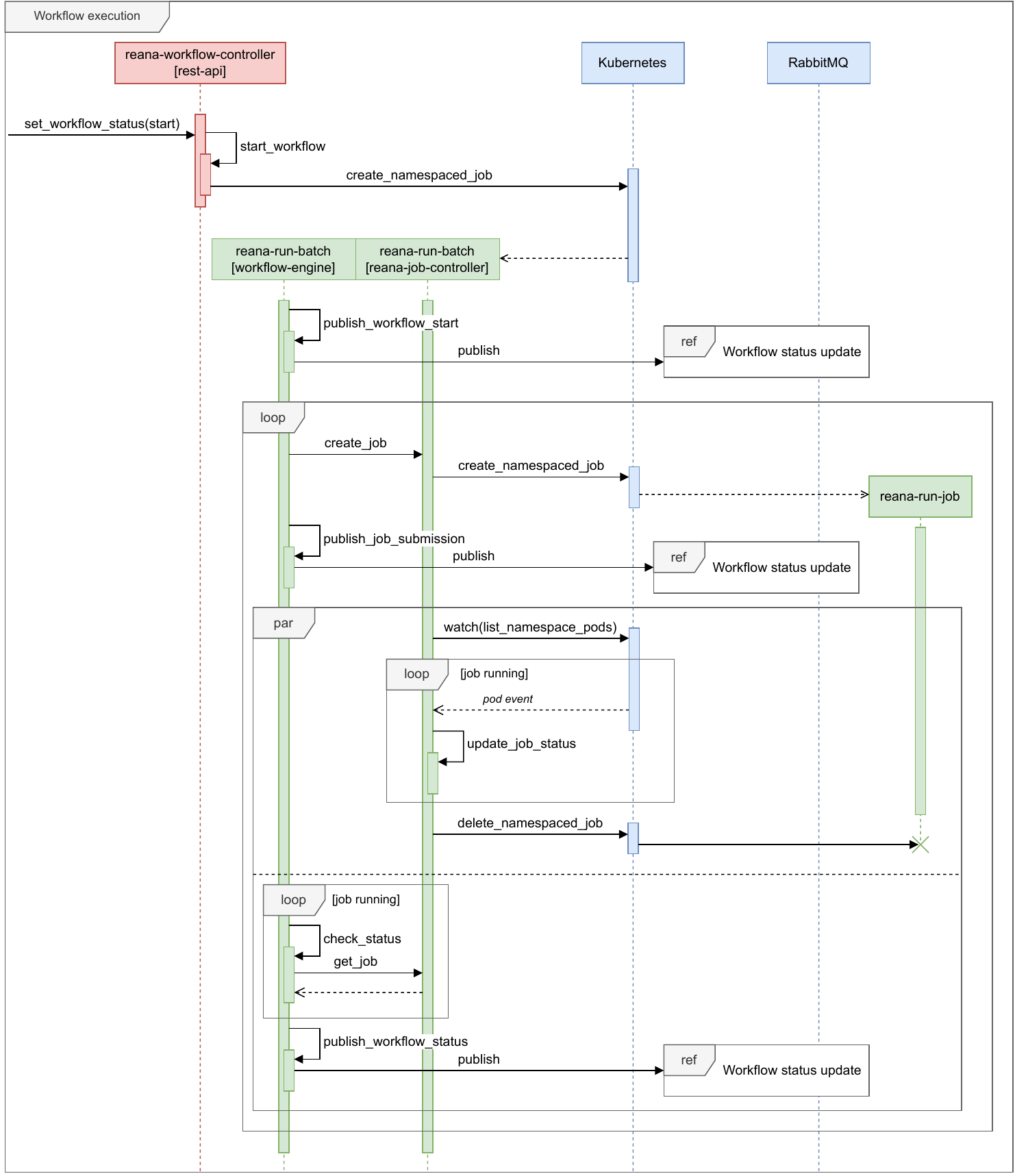}
\caption{The sequence diagram showing how the REANA executes scheduled workflows.
Note the interplay between the scheduler and the Kubernetes cluster.
The pod creation offers another space for optimisations.
The workflow execution status monitoring is carried out by a watching loop.
The workflow jobs are started for each workflow step.
The termination procedures are further illustrated in Figure~\ref{fig:reanascheduler4}.}
\label{fig:reanascheduler3}
\end{figure}

Figure~\ref{fig:reanascheduler3} shows the stage of the running of the workflow after it has been accepted for execution.
Note the interplay of the REANA platform with the underlying Kubernetes cluster: the job is scheduled using the Kubernetes native job scheduler mechanism which include additional scheduling delays that needed to be taken into account for optimisation.
The progress of the workflow is monitored until the workflow execution terminates.
The workflow steps are launched when the worker nodes are free to run the workload.
The status of jobs is published in the message queue.

\begin{figure}
\centering
\includegraphics[width=0.9\textwidth]{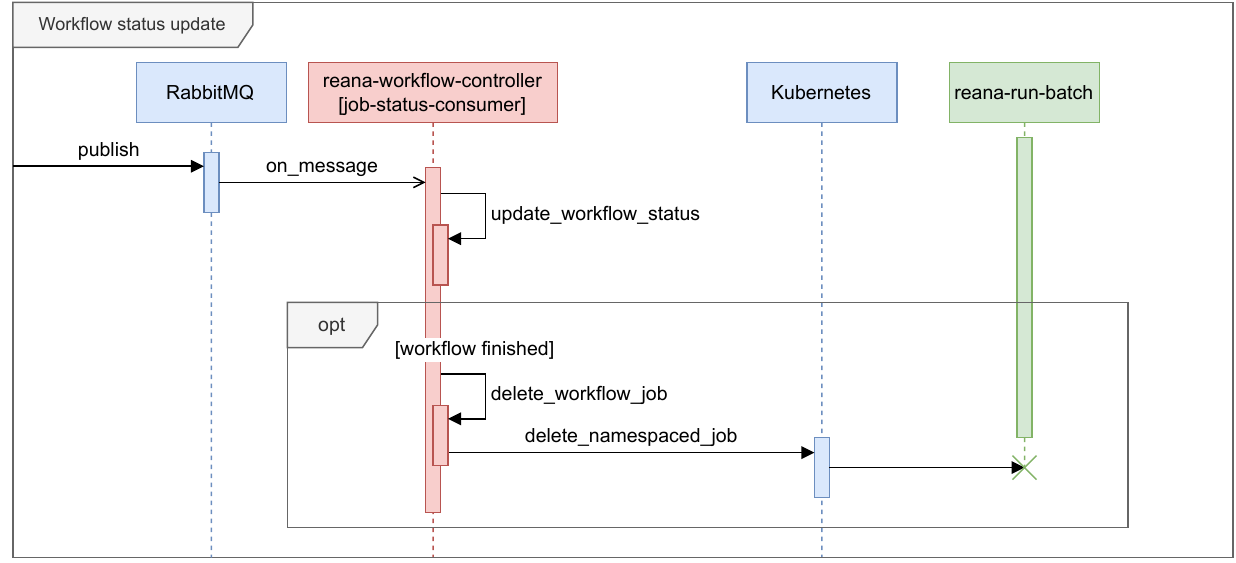}
\caption{The sequence diagram showing how REANA updates workflow statuses and terminates finished workflows.
The procedure involves consuming the message queue, closing the Kubernetes pods, and updating the database about the status of the workflow run.
In case of launching several thousands of concurrent workflows, these processes also have to be optimised.}
\label{fig:reanascheduler4}
\end{figure}

Figure~\ref{fig:reanascheduler4} shows the termination stage of the workflow.
When all the steps are finished and the results are produced, the system has to delete the Kubernetes pod and update the status of the workflow in both the message queue and the database.
This constituted another layer of optimisations in order to handle any status handling processes in an asynchronous manner whilst the platform is starting the new incoming workflows.

\section{Results}\label{sec:results}

We have improved the REANA platform scheduling performance in order to maximise the scheduling throughput of incoming workflows at the various stages of the workflow life cycle as described in Section~\ref{sec:method}.
A special attention was paid to measure the CPU and Memory usage of the cluster nodes.

Figure~\ref{fig:testnodes} shows a typical snapshot of the status of cluster nodes running the pMSSM workloads.
We have used nodes of the \texttt{m2.xlarge} flavour which consist of 16 GiB of available memory and 8 virtual cores.
One can see the efficient use of cores of the cluster resulting from tuning REANA parameters such as the number of nodes running workflow orchestration tasks, the number of nodes running the pMSSM workflow step jobs themselves, as well as the memory request limits for each ntupling job of the first pMSSM workflow stages.

\begin{figure}
\centering
\footnotesize
\begin{verbatim}
    $ kubectl top nodes
    NAME                                CPU(cores)   CPU%   MEMORY(bytes)   MEMORY%
    reanaatlas1-3slyowp42qex-node-15    7858m        98%    12033Mi         82%
    reanaatlas1-3slyowp42qex-node-16    7848m        98%    12083Mi         83%
    reanaatlas1-3slyowp42qex-node-17    7846m        98%    12210Mi         83%
    reanaatlas1-3slyowp42qex-node-18    7773m        97%    8995Mi          61%
    reanaatlas1-3slyowp42qex-node-19    7864m        98%    11516Mi         79%
    reanaatlas1-3slyowp42qex-node-20    7843m        98%    12177Mi         83%
    reanaatlas1-3slyowp42qex-node-21    7376m        92%    8698Mi          59%
    reanaatlas1-3slyowp42qex-node-22    7817m        97%    11201Mi         77%
    reanaatlas1-3slyowp42qex-node-23    7748m        96%    9978Mi          68%
    reanaatlas1-3slyowp42qex-node-24    7854m        98%    12161Mi         83%
    reanaatlas1-3slyowp42qex-node-25    7868m        98%    12293Mi         84%
    reanaatlas1-3slyowp42qex-node-26    7787m        97%    10991Mi         75%
\end{verbatim}
\caption{An example of the benchmark tests running in the CERN Computer Centre.
The REANA scheduling parameters were optimised to maximise the CPU utilisation and the Memory consumption on the cluster for the typical pMSSM ntupling job parallelism (see Figure~\ref{fig:dag}).
Note the very good efficiency of CPU cores in the above screenshot.}
\label{fig:testnodes}
\end{figure}

\begin{figure}
\centering
\includegraphics[width=0.45\textwidth]{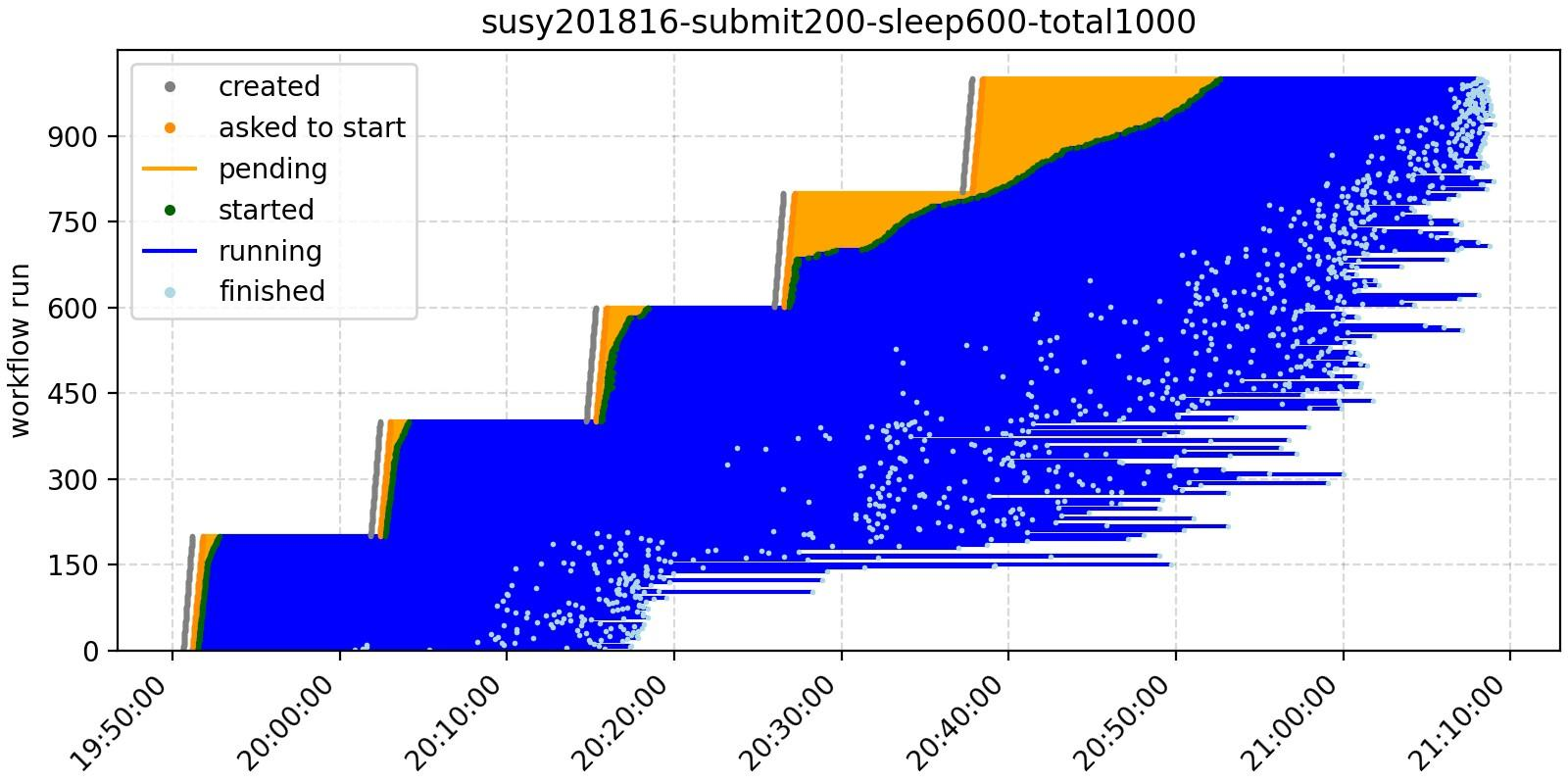}
\includegraphics[width=0.45\textwidth]{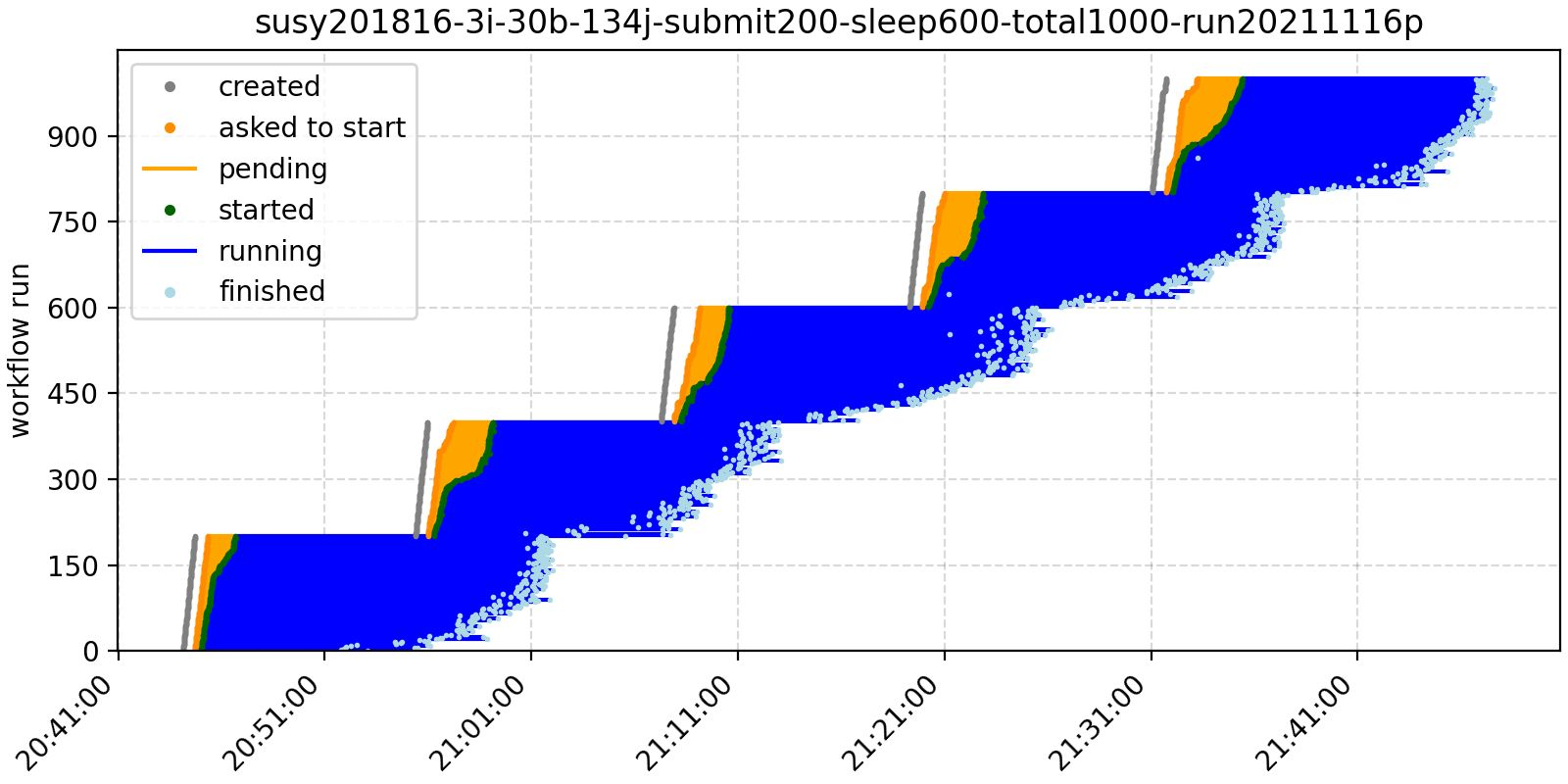}
\caption{A scalability test submitting 200 workflows every 10 minutes.
A cluster with 448 cores (left) cannot keep up with the load.
A cluster with 1072 cores (right) can comfortably hold the incoming workload.}
\label{fig:testresults}
\end{figure}

Figure~\ref{fig:testresults} shows the results of one of our scalability experiment that consisted of submitting 200 new pMSSM workflows every 10 minutes.
A cluster with 448 cores presented on the left cannot keep up with such a workload: note the increasing scheduling waiting times (plotted in the orange colour) as well as increasing workflow run times (plotted in blue).
The overflow happens because the cluster is allowing more workflows than it can hold.
However, note how the same cluster with 1072 cores presented on the right of the Figure holds the same workload very comfortably.

\begin{figure}
\centering
\includegraphics[width=0.9\textwidth]{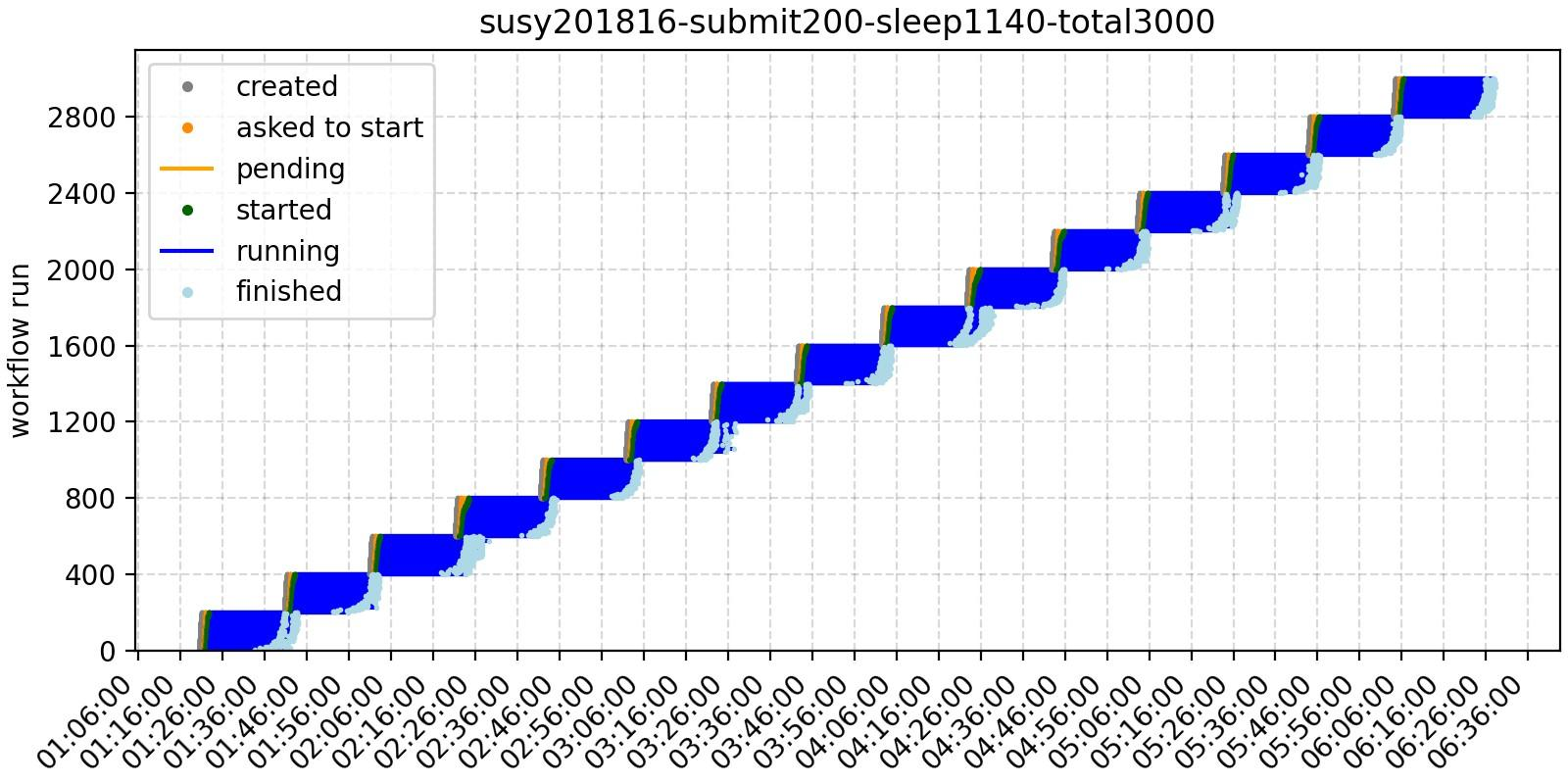}
\caption{The workload burndown throughput rate is sustainable over a long period of time.}
\label{fig:testresultslong}
\end{figure}

Figure~\ref{fig:testresultslong} shows the same kind of experiment executed over a longer period of time.
This helped to ensure that the platform can sustain the constantly increasing stream of incoming workloads.

We have run several benchmarking experiments in the CERN Computer Centre and, to test the portability, performed a few runs also on the Google Cloud Platform.
This allowed to prove the applicability of the approach on various compute backends, facilitating future reproducibility of containerised workflows irrespective of their original computing environments.

\section{Conclusions}\label{sec:conclusions}

ATLAS searches for new physics are being effectively preserved together with containerised computational workflow recipes as part of the ATLAS RECAST project.
This enables their future reuse and reinterpretation and greatly facilitates the running of efficient pMSSM studies over a large collection of individual analyses.

We have launched several ATLAS pMSSM workflows on the REANA reproducible analysis platform and studied the performance from workflow scheduling up to workflow execution and termination procedures with the aim of allowing running several thousands of these workflows to cover a sufficient number of pMSSM model points.

The REANA platform has been internally optimised to allow faster workflow scheduling, processing and terminating procedures on an individual workflow level as well as under the stressing conditions of processing many incoming concurrent workloads.
A set of benchmarking experiments allowed to optimise and tune the REANA system for the pMSSM workloads on the Kuberentes clusters ranging from medium to large sizes (from 500 to 5000 cores).
It was essential to adjust REANA scheduling parameters to the type of the pMSSM workload in order to ensure the best throughput and the efficient cluster CPU and memory resource utilisation.

The developed system was tested on the CERN Computer Centre as well as on the Google Cloud Platform in order to ensure the reproducibility of the approach and is fully ready to run large-scale ATLAS pMSSM reinterpretations of LHC Run-2 analyses.
The first results by the ATLAS collaborations are being published~\cite{ATLAS:2023oun}.

\section*{Acknowledgements}\label{sec:acknowledgements}

L.H. is supported by the Excellence Cluster ORIGINS, which is funded by the Deutsche Forschungsgemeinschaft (DFG, German Research Foundation) under Germany's Excellence Strategy - EXC-2094-390783311.

M.F. is supported by the U.S. National Science Foundation (NSF) under Cooperative Agreement OAC-1836650 (IRIS-HEP).

% Don't use \bibliographystyle
% the style is already called in woc.bst, so ensure it is in the top level directory
% \bibliography{bib/ref}  # Reproducing typesetting of accepted proceedings, so not using bibliography

\end{document}